\documentclass[aps,prl,reprint,superscriptaddress]{revtex4-1}

\usepackage{bbm}
\usepackage{amsmath,leftidx}
\usepackage{graphicx}
\usepackage{times}
\usepackage{CJK}
\usepackage{color}
\usepackage[normalem]{ulem}

\begin{document}


\begin{CJK*}{UTF8}{}
\title{Crossover of Correlation Functions near a Quantum Impurity in a Tomonaga-Luttinger Liquid}
 \CJKfamily{bsmi}

\author{Chung-Yu Lo (羅中佑)}
\affiliation{ Department of Physics, National Tsing Hua University, Hsinchu 30013, Taiwan}

\author{Yoshiki Fukusumi (福住吉喜)}
\affiliation{Institute for Solid State Physics, University of Tokyo, Kashiwa 277-8581, Japan}


\author{Masaki Oshikawa (押川正毅)}
\affiliation{Institute for Solid State Physics, University of Tokyo, Kashiwa 277-8581, Japan}

\author{Ying-Jer Kao (高英哲)}
\email{yjkao@phys.ntu.edu.tw}
\affiliation{ Department of Physics, National Taiwan University, Taipei 10607, Taiwan}
\affiliation{ National Center for Theoretical Sciences, Hsinchu 30013, Taiwan}

\author{Pochung Chen (陳柏中)}
\email{pcchen@phys.nthu.edu.tw}
\affiliation{ Department of Physics, National Tsing Hua University, Hsinchu 30013, Taiwan}
\affiliation{ National Center for Theoretical Sciences, Hsinchu 30013, Taiwan}

\date{\today}

\begin{abstract}
An impurity in a Tomonaga-Luttinger liquid leads to a crossover
between short- and long-distance regime which describes many physical
phenomena. However, calculation of the entire crossover of
correlation functions over different length scales has been difficult.
We develop a powerful numerical method based on infinite DMRG
utilizing a finite system with infinite boundary conditions,
which can be applied to correlation functions near an impurity.
For the $S=1/2$ chain, we demonstrate that the full crossover
can be precisely obtained, and that their limiting behaviors
show a good agreement with field-theory predictions.
\end{abstract}


\maketitle
\end{CJK*}

In one-dimensional (1D) systems, even weak Coulomb interactions have dramatic effects and the Fermi liquid theory describing their higher dimensional counterparts breaks down. This results in a Tomonaga-Luttinger liquid (TLL)~\cite{Tomonaga:1950hq,*Luttinger:1963pt,Giamarchi:2004ix}, which is nothing but a relativistic free boson field theory.
The TLL  behaviors have  been experimentally demonstrated in carbon nanowires~\cite{Laroche:2014qf,*Ishii:2003vn,*Yao:1999fj,*Kim:2007gf,*Postma:2000ul}, allowing for further studies of the electron transport in 1D quantum wires.  On the theory side,  there exists a plethora of powerful analytical and numerical methods available to study the behavior of 1D systems. Analytical tools such as the bosonization, conformal field theory (CFT) and the renormalization group (RG) can be employed to analyze the physical properties of TLL~\cite{Giamarchi:2004ix,Cardy:2010zl,Affleck:2010cr}.

An important class of problems is the effects of a quantum impurity on a TLL~\cite{Kane:1992kx,*Kane:1992gj,Furusaki:1993tw,Matveev:1993xq,Wong:1994zt,Rylands:2016eu,Shi:2016rm}. 
In the simplest setting, Kane and Fisher have shown that  a single quantum impurity affects the transport property of the TLL in an essential way~\cite{Kane:1992kx,*Kane:1992gj,Furusaki:1993tw}: when the interaction is attractive the system renormalizes to a fixed point corresponding to a single fully connected wire. When the interaction is repulsive, however, the system  renormalizes to two  disconnected wires. An equivalent problem was also studied in a context of quantum spin chains~\cite{EggertAffleck1992}.
In terms of CFT, a RG fixed point of the impurity problem is associated to a conformally invariant boundary condition (CIBC)~\cite{Saleur-Lecture1998}.
Thus the first question in the impurity problem is the classification of CIBCs.
While nontrivial CIBCs appear in various settings~\cite{Affleck:1991pz,Oshikawa:2006sw}, only the 
simple Dirichlet and Neumann boundary conditions of the free boson field theory
are relevant for 
the original Kane-Fisher problem (a single quantum impurity) in
a spinless single-channel interacting TLL.
For each CIBCs, correlation functions can be calculated with boundary CFT techniques.
However, the system is renormalized to the low-energy/large-distance (infrared, IR) fixed point only asymptotically.
In order to describe various observable properties, such as finite-temperature properties, we need to describe the RG flow towards the IR fixed point, not just the CIBC corresponding to the IR fixed point. 
The system is often renormalized close to a high-energy/short-distance (ultraviolet, UV) fixed point first, before flowing towards the IR fixed point. In such a case, the finite-energy/finite-distance properties can be described as a crossover between the UV and IR fixed points. 
The crossover phenomena cannot be dealt with the boundary CFT techniques alone. In some cases, the crossover of a physical quantity can be exactly obtained in terms of an integrable boundary RG flow~\cite{FendleyLudwigSaleur}.
Nevertheless, for more general quantities, and for other settings, numerical approach is indispensable to describe the crossover.

In general it is difficult to simulate 1D (boundary) critical systems, of which the TLL is an example, because large system sizes are required to capture the asymptotic behavior. Lo \textit{et al.}\cite{Lo:2014vr} use a scale-invariant tensor network to directly extract scaling operators and scaling dimensions for both bulk and boundary CFTs. Although the method can successfully describe the physics at the IR fixed point, it can not probe the UV to IR RG flow. Rahmani \textit{ et al.}~\cite{Rahmani:2010jr,Rahmani:2012bq} perform a conformal mapping of the wire junction to a finite strip so that a finite-size DMRG calculation can be carried out. However,  an {\em ad hoc} mirror boundary condition has to be added. 
Furthermore, the conformal mapping makes it necessary to use of the chord distance, instead of the direct site distance.   It is therefore difficult to probe short-distance and crossover behavior using this approach. An improved numerical method is hence called for.

In this Letter, we present a numerical method based on an infinite DMRG (iDMRG) scheme that allows us to directly simulate the junction of semi-infinite TLL wires via infinite boundary condition (IBC) and study the crossover from the impurity site to the long length scale. We are able to obtain various correlation functions which are in agreement, at both short and long length scale,  with those obtained by the boundary perturbation theory based on bosonization.\cite{Fateev:1997nn, Fendley:1998pq, Fendley:1998sq, PhysRevLett.75.4492, PhysRevB.58.5529}

\begin{figure}[bt]
  \includegraphics[width=0.9\columnwidth]{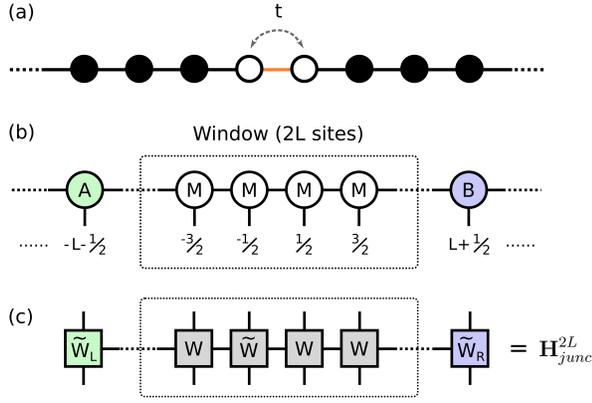}
  \caption{(Color online) (a) Sketch of the junction with a link of strength $t$. (b) Infinite matrix product state diagram for $|\Psi\rangle_{\text{junc}}^{G, 2L}$. (c) Matrix product operator diagram for $H_{\text{junc}}^{2L}$.}
  \label{fig:sketch}
\end{figure}

We start from two semi-infinite wires of spinless electrons. The two wires are connected by a link of strength $t$ to form the junction as sketched in Fig.\ref{fig:sketch}(a). Using Jordan-Wigner transformation, the wire and the link Hamiltonians can be written in spin language as
\begin{equation}
  H_{\text{wires}} = \sum_{\substack{\mu=\alpha,\beta,\\ i \in \mathcal{Z}^+ + \frac{1}{2} }}
   -\left(  S^{+\mu}_i S^{- \mu}_{i+1} +S^{- \mu}_i S^{+ \mu }_{i+1}  \right) + V S^{z \mu}_i S^{z \mu}_{i+1},
\end{equation}
and
\begin{equation}
 H_{\text{link}} = -t ( S^{+\alpha}_{1/2} S^{-\beta}_{1/2} +S^{-\alpha}_{1/2} S^{+\beta}_{1/2} ),
\label{eq.link}
\end{equation}
respectively, where $\alpha$, $\beta$ are wire indices. The junction Hamiltonian is then defined as $H_{\text{junc}}= H_{\text{wires}}+H_{\text{link}}$. We also define a bulk Hamiltonian for a single infinite TLL wire as
\begin{equation}
  H_{\text{bulk}} =  \sum_{i \in \mathcal{Z} + \frac{1}{2} }
  -\left( S^x_i S^x_{i+1} + S^y_i S^y_{i+1} \right) + V S^z_i S^z_{i+1}.
  \label{eq:bulk}
\end{equation}
It differs from a junction with strength $t=1$ only by the interaction across the junction: $V S^{z\alpha}_{1/2} S^{z\beta}_{1/2}$. Both the semi-infinite wires of the junction and the bulk wire are described by the TLL  theory with Luttinger parameter $g=\pi/(2\arccos(-V/2))$. We will consider three inter-wire correlation functions: $\langle S^{\alpha + }_i S^{\beta -}_i \rangle$, $\langle S^{\alpha z }_i S^{\beta z}_i \rangle$, and $\langle J^{\alpha}_i J^{\beta}_i \rangle$. Here the current operator is defined as $J^\mu_i \equiv -i ( S^{\mu +r}_{i-1/2} S^{\mu -}_{i+1/2}  - S^{\mu +}_{i+1/2} S^{\mu -}_{i-1/2} )$.

To find the ground state of $H_\text{junc}$, we use the IBC~\cite{Michel:2010uo, *Phien:2012dr, *Phien:2013kh} to construct an effective junction Hamiltonian for a finite-size window that contain the impurity. The effective Hamiltonian is then used to optimized the ground state within the window. This results in an optimized ground state in the form of an infinite matrix product state (iMPS), from which correlation functions within the window can be easily calculated. Specifically we start from the translationally invariant bulk Hamiltonian $H_{\text{bulk}}$. We assume that its ground state is described by an one-site (or two-site) translationally invariant iMPS:
\begin{align}
  |\Psi\rangle_{\text{bulk}}^{G}& = \sum_{ \{s_i\} }
  \cdots \lambda \Gamma^{s_{i-1}} \lambda \Gamma^{s_i} \lambda \Gamma^{s_{i+1}} \lambda \Gamma^{s_{i+2}} \cdots
  |\mathbf{s}\rangle,\nonumber\\
 &= \sum_{s_i} \cdots A^{s_{i-1}} A^{s_i} \lambda B^{s_i} B^{s_{i+1}} \cdots |\mathbf{s}\rangle,
\end{align}
where $ |\mathbf{s}\rangle=|\dots s_{i-1}, s_i, s_{i+1}, s_{i+2} \dots \rangle$ and $s_i$ are local spin basis. Furthermore, $\Gamma^{s}=\Gamma$ are site-independent $d\times D\times D$ tensors and $\lambda$ is a $D\times D$ diagonal matrix, where $d$ and $D$ are physical dimension and bond dimension respectively. The second line corresponds to the mixed canonical form with $A^s=A=\lambda \Gamma$ and $B^s=B=\Gamma \lambda$. Here $A$ and $B$ satisfy the left and right canonical form constraints respectively. They can be obtained by optimizing $ |\Psi\rangle_{\text{bulk}}^{G}$ with the bulk Hamiltonian $H_{\text{bulk}}$ via any conventional iDMRG algorithm. Due to the presence of the quantum impurity, the translational invariance is broken and  a translational invariant iMPS is  no longer a good ansatz for the ground state of $H_{\text{junc}}$. On the other hand, since $H_{\text{junct}}$ differs from $H_{\text{bulk}}$ only at the impurity sites, we expect that far away from these sites the ground states of  $H_{\text{junct}}$ and $H_{\text{bulk}}$ should resemble each other locally. We hence assume that there is a finite window of size $2L$ with sites $ i \in [-L-\frac{1}{2}, \cdots, L+\frac{1}{2} ]$  within which the ground states of $H_{\text{junc}}$ and  $H_{\text{bulk}}$ differ, while outside this window they are locally described by the same matrices. This leads to the following iMPS ansatz for the ground state of $H_{\text{junc }}$:
\begin{equation}
  |\Psi\rangle_{\text{junc}}^{G, 2L} =
  \sum_{s_i} \cdots A^{s_{-L-1}} \left ] \right [ M^{s_{-L}} \cdots M^{s_L} \left ] \right [ B^{s_{L+1}} \cdots |\mathbf{s}\rangle,
\end{equation}
as sketched in Fig.\ref{fig:sketch}(b). Here $L$ is an adjustable parameter that can be easily enlarged. And the $M$ matrices are optimized with an effective Hamiltonian as described below.

Starting from $H_{\text{bulk}}$ in the form of matrix product operators (MPOs) \cite{Schollwock:2011gl, *McCulloch:2007gi}:
\begin{equation}
  H_{\text{bulk}} = \cdots W_{-L-\frac{3}{2}} W_{-L-\frac{1}{2}} \cdots W_{-\frac{1}{2}} W_{\frac{1}{2}} \cdots W_{L+\frac{1}{2}} W_{L+\frac{3}{2}} \cdots
\end{equation}
where the matrix $W_i=W$ is site independent. The effective Hamiltonian of the finite window can be expressed as
\begin{equation}
  H_{\text{junc}}^{2L} = \widetilde{W}_{\mathcal{L}} W_{-L-\frac{1}{2}} \cdots \widetilde{W}_{-\frac{1}{2}} W_{\frac{1}{2}} \cdots W_{L+\frac{1}{2}} \widetilde{W}_{\mathcal{R}}
\end{equation}
as sketched in  Fig.\ref{fig:sketch}(c). Here $W_{-1/2}$ are replaced by $\widetilde{W}_{-1/2}$ to represent $H_{\text{link}}$.
Furthermore, the left and right IBCs, $\widetilde{W}_{\mathcal{L}}$ and $\widetilde{W}_{\mathcal{R}}$, are constructed from the left and right dominant eigenvectors of the generalized transfer matrices   $T_\mathcal{L} = \sum_{ss^\prime} \langle s | W | s^\prime\rangle A^{s^\prime \dagger} A^{s}$ and  $T_\mathcal{R} = \sum_{ss^\prime} \langle s | W | s^\prime\rangle B^{s^\prime} B^{s \dagger}$ respectively \cite{Phien:2012dr}. Here the IBCs are used to represent the semi-infinite extensions of the bulk system to the left and right. In this way, we reduce an infinite-size system to an effective finite-size one~\cite{SM}. Once $H_{\text{junc}}^{2L}$ is obtained, one can use any conventional finite-size MPS/DMRG algorithm to optimize the $M$ matrices. Also, with the left and right-canonical conditions satisfied by $A$ and $B$ matrices, correlation functions within the window can be calculated using only $M$ matrices.

\begin{table}[tb]
\caption{Dominant exponents for each correlation function}
\begin{center}
\begin{tabular}{|c|c|c|c|c|c|c|}
\hline
 & \multicolumn{3}{|c|}{$g>1$} & \multicolumn{3}{|c|}{$g<1$} \\
\hline
 & Bulk IR & IR & UV & Bulk IR & IR & UV \\
\hline
  $\langle S^{+,\alpha }_i S^{-,\beta}_i \rangle$ & $\frac{1}{2g}$ & $\frac{1}{2g}$ & $\frac{3}{2g}-1$ &  $\frac{1}{2g}$ & $\frac{3}{2g}-1$ & $\frac{1}{2g}$ \\
\hline
  $\langle S^{z,\alpha }_i S^{z,\beta}_i \rangle$ & 2 & 2 & $\frac{2}{g}$ & $2g$ & $2g+\frac{2}{g}-2$ & $2g$ \\
\hline
  $\langle J^{\alpha}_i J^{\beta}_i \rangle$  & 2 & 2 & $\frac{2}{g}$ & $2$ & $\frac{2}{g}$ & $2$ \\
\hline
\end{tabular}
\end{center}
\label{table:exponents}
\end{table}

\begin{figure}[tb]
  \includegraphics[width=0.9\columnwidth]{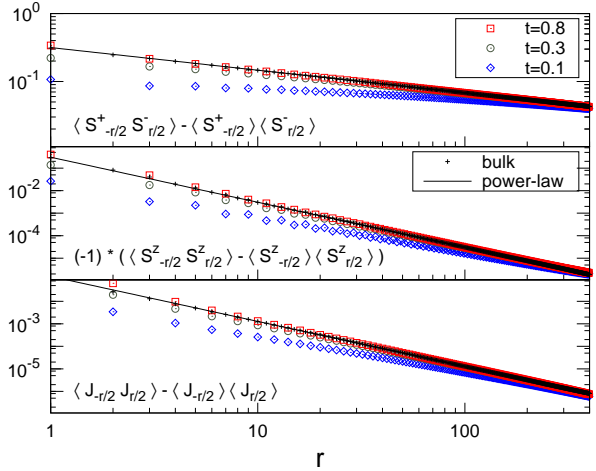}
  \caption{(Color online) $\langle S^{\alpha, + }_i S^{\beta, -}_i \rangle$, $\langle S^{\alpha, z }_i S^{\beta, z}_i \rangle$, and $\langle J^{\alpha}_i J^{\beta}_i \rangle$ correlation functions for $g=1.5$. Data for the bulk and junctions with $t=0.1, 0.3, 0.8$ are plotted. Solid lines are power law fitting to the bulk data with bulk exponents from bosonization. (Cf. Table.\ref{table:exponents}.)}
  \label{fig:g>_IR}
\end{figure}

When $g>1$,  at long distance (IR limit) the system is renormalized to a single wire with the same Luttinger parameter. For a weaker link (smaller $t$),  we expect that it would take longer distance for the system to heal from the perturbation due to the impurity. Figure~\ref{fig:g>_IR} shows spin-spin and current-current correlation functions between two leads: $\langle S^{+,\alpha }_i S^{-,\beta}_i \rangle$, $\langle S^{z,\alpha }_i S^{z,\beta}_i \rangle$, and $\langle J^{\alpha}_i J^{\beta}_i \rangle$ for junctions with $g=1.5$ and various $t$ as well as a bulk wire. We observe that at IR limit all correlation functions merge into their  bulk counterparts. This confirms that asymptotically the system is renormalized to a single defect-free wire.

\begin{figure}[tb]
  \includegraphics[width=0.9\columnwidth]{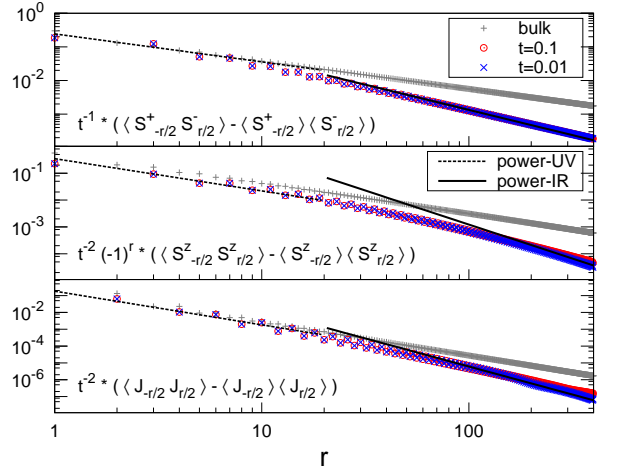}
  \caption{(Color online) Rescaled $\langle S^{\alpha, + }_i S^{\beta, -}_i \rangle$, $\langle S^{\alpha, z }_i S^{\beta, z}_i \rangle$, and $\langle J^{\alpha}_i J^{\beta}_i \rangle$ correlation functions for $g=0.6$. Data for the bulk and junctions with $t=0.1, 0.01$ are plotted. Solid (dotted) lines are power law fitting to the long (short) distance data with IR (UV) exponents from bosonization.  (Cf. Table.\ref{table:exponents}.) }
  \label{fig:g<}
\end{figure}

When $g<1$, in contrast, we expect that in the  IR limit the system is renormalized into two disconnected semi-infinite wires. For non-zero $t$, we still expect that the correlation functions to decay as a power law, but with exponents that are larger than the bulk counterparts. In Fig.\ref{fig:g<}, we plot the same correlation functions for $g=0.6$ and $t=0.1, 0.01$ as well as the single bulk wire. A scaling prefactor of $t^{-1}$ or $t^{-2}$ is also included in order to collapse the curves with different $t$'s. It is clear that the correlators in the IR limit decay faster than their bulk counterparts, supporting the picture that at IR limit the system is renormalized into broken wires.

To further understand the behavior of these correlation functions in
both the UV and the IR limits, we use boundary perturbation theory to
determine the exponents of the power laws. To the leading order, we
derive the exponents of uniform and staggered part of the correlation
functions respectively.
In the bosonization framework, the system is described by the TLL
with the Lagrangean density\begin{equation}
 \mathcal{L} = \frac{1}{2\pi g} (\partial_\mu \phi)^2 .
\end{equation}
In the leading orders,
the spin and current operators are expressed as~\cite{Giamarchi:2004ix,Lukyanov:2002fg}
\begin{align}
  S^{-}_j & \sim   e^{- i \theta }\left(b +c\left(-1\right)^{j} \cos{(2\phi)}) \right) , \nonumber\\
  S^z_j & \sim - \frac{1}{\pi}\frac{\partial\phi}{\partial x}+a(-1)^j \sin{(2\phi)} , \nonumber\\
  J_j & =i\left(S^{+}_{j+1}S^{-}_{j}-S^{+}_{j}S^{-}_{j+1}\right) \sim \frac{gv}{\pi} \frac{\partial \theta}{\partial x},
\end{align}
where $a$, $b$, and $c$ are constants.
Here $\theta$ is the dual field of $\phi$ defined by 
$\theta \equiv \frac{1}{g} (\phi_L - \phi_R)$, where
$\phi = \phi_L(\bar{z}) + \phi_R(z)$ is the chiral decomposition
into left/right-movers with the complex coordinate $z=x+it$.
The correlation functions of the chiral fields on the full complex plane
(without an impurity) read
\begin{align}
 \langle \phi_R(0) \phi_R(z) \rangle & \sim
  - \frac{g}{4} \log{z} + \mbox{const.} , \nonumber\\
 \langle \phi_L(0) \phi_L(\bar{z}) \rangle & \sim
  - \frac{g}{4} \log{\bar{z}} + \mbox{const.}.
\end{align}
In this convention, $\phi$ and $\theta$ are compactified respectively
as $\phi \sim \phi + \pi$ and $\theta \sim \theta + 2\pi$.

The geometry is half plane with the interaction on the line $x=0$.
In the limit $t=0$, the system is two decoupled half-chains. The end of each half-chain is renormalized to the Dirichlet boundary condition $\phi^\gamma=0$ for $\gamma=\alpha,\beta$.
We then introduce the link~Eq.~\eqref{eq.link} between the two decoupled chains through the boundary.
Let us consider the correlation function
$\left\langle S^{\alpha + }_{i + \frac{1}{2}} S^{\beta -}_{i + \frac{1}{2}} \right\rangle$, across the link. Obviously it vanishes when $t=0$.
In the first order of $t$, the correlation function is given as
\begin{equation}
   t \int d\tau \;
  \left[ \frac{1}{2}
  \langle S^{+\alpha}_{i+\frac{1}{2}}(0) S^{\alpha-}_{\frac{1}{2}}(\tau)\rangle_{D}
  \langle S^{\beta +}_{\frac{1}{2}}(\tau) S^{-\beta}_{i+\frac{1}{2}}(0) \rangle_{D}
\right].
\label{eq.SpSm_1st}
\end{equation}
The problem is thus reduced to calculation of the correlation functions with the Dirichlet boundary condition~\cite{PhysRevB.58.5529}.
The Dirichlet boundary condition can be solved by an analytical continuation of $\phi^{\gamma,R}$ to $x<0$ and $\phi^{\gamma,L}(x,\tau) \equiv - \phi^{\gamma,R}(-x,\tau).$
Using this, the correlation function is given as
\begin{equation}
  \label{eq:SS_D}
  \left\langle S^{\alpha + }_{i + \frac{1}{2}} S^{\beta -}_{i + \frac{1}{2}}\right \rangle  = t \left[
  C_0 r^{ -\left(\frac{3}{2g} -1 \right) } +
  C^\prime_0 (-1)^{r} r^{ -\left(\frac{3}{2g} +2g-1 \right) } \right],
\end{equation}
where $i \in \mathcal{Z}$, $r=2i+1$, and $C_0, C^\prime_0$ are constants. Similarly we find
\begin{equation}
  \label{eq:ZZ_D}
  \left\langle S^{\alpha z }_{i + \frac{1}{2}} S^{\beta z}_{i + \frac{1}{2}} \right\rangle   = t^2 \left[
  C_0 r^{ -\left( \frac{2}{g} \right) } +
  C^\prime_0 (-1)^{r}  r^{ -\left( 2g + \frac{2}{g} -2 \right) } \right],
\end{equation}
where $r=2i+1$ and
\begin{equation}
  \label{eq:JJ_D}
  \left\langle J^{\alpha}_i J^{\beta}_i \right\rangle = t^2 C_0 r^{-\frac{2}{g} },
\end{equation}
where $r=2i$. These results describe the IR behavior for $g<1$ and the UV behavior for $g>1$. We find that for $g>1$, the uniform part always dominates and the UV exponents are $3/2g-1$, $2/g$, and $2/g$ respectively. In contrast, for $g<1$ the staggered part of the  $\langle S^{z,\alpha }_i S^{z,\beta}_i \rangle$ becomes dominant and the IR exponents become $3/2g-1$, $2g+2/g-2$, and $2/g$ respectively.

The case of a weak barrier (small $1-t$), on the other hand, corresponds to the free boundary condition. For this case we regard the junction as a defect in CFT with
\begin{equation}
  H_{\text{barrier}}
  = (1-t) ( S^{\alpha +}_{1/2} S^{\beta -}_{1/2} +S^{\alpha -}_{1/2} S^{\beta +}_{1/2} ) - V S^{z\alpha}_{1/2} S^{z\beta}_{1/2}.
\end{equation}
We evaluate this defect by using operator product expansion for CFT. By the usual perturbation theory, we find
\begin{equation}
  \label{eq:SS}
  \left\langle S^{\alpha + }_{i + \frac{1}{2}} S^{\beta -}_{i + \frac{1}{2}}\right \rangle =
  C_0 r^{ -\frac{1}{2g} } +  C^\prime_0 (-1)^r r^{ -\frac{1}{2g}+2g },
\end{equation}
\begin{equation}
  \label{eq:ZZ}
  \left\langle S^{\alpha z }_{i + \frac{1}{2}} S^{\beta z}_{i + \frac{1}{2}}\right \rangle  =
  C_0 r^{ -2 } + C^\prime_0 r^{ -2g },
\end{equation}
where $i \in \mathcal{Z}$ and $ r=2i+1 $.
\begin{equation}
  \label{eq:JJ}
  \left\langle J^{\alpha}_i J^{\beta}_i \right\rangle  = C_0 r^{ -2 }
\end{equation}
where $ r=2i $. These results describe the IR behavior for $g>1$ and the UV behavior for $g<1$. We find that for $g>1$, the uniform part always dominate and the IR exponents are $1/2g$, $2$, and $2$ respectively. In contrast,  for $g<1$ the staggered part of the $\langle S^zS^z\rangle$  dominates and the UV exponents are $1/2g$, $2g$, and $2$. These results also describe the IR behavior for a single bulk TLL wire. In Table~\ref{table:exponents}, we summarize the dominant exponent for each correlation function~\cite{SM}.

We now compare the numerical results against these power laws. Figure~\ref{fig:g>_IR} shows the fit of the bulk correlation functions to Eqs.~(\ref{eq:SS}), (\ref{eq:ZZ}), and (\ref{eq:JJ}) for the case of an attractive interaction $g=1.5$. The numerical results confirm that in the IR limit all correlation functions decay with corresponding bulk exponents. Also, we see that for a weaker link (smaller $t$), it takes longer distance for the system to reach the IR limit, indicating it needs to more steps to renormalize away the impurity.

In Fig.~\ref{fig:g<} we compare the numerical results against the bosonization results for the case of a repulsive interaction $g=0.6$. At the IR limit the system is renormalized into two disconnected semi-infinite wires, which corresponds to the Dirichlet boundary condition. Very weak links with $t=0.1, 0.01$ are used to probe the IR behavior. We observe at long distance not only the exponents of the rescaled correlation function agree with Eqs.~(\ref{eq:SS_D}), (\ref{eq:ZZ_D}), and (\ref{eq:JJ_D}), but also $t$-dependent prefactors agree. This suggests that the perturbative calculation indeed captures the correct physics of the junction. On the other hand, while small $1-t$ is assumed in the derivation of Eqs.~(\ref{eq:SS}), (\ref{eq:ZZ}), and (\ref{eq:JJ}), the exponents describe well the numerical results at short distances even when $t$ is small (and thus $1-t$ is not small).
Furthermore the same scaling prefactors also result in data collapse at short distance. Interestingly, this indicates that near the junction, the system does not know which fixed point it should renormalize into, and the correlation in the short distance looks like a bulk wire, scaled with the junction strength $t$.

\begin{figure}[tb]
  \includegraphics[width=0.9\columnwidth]{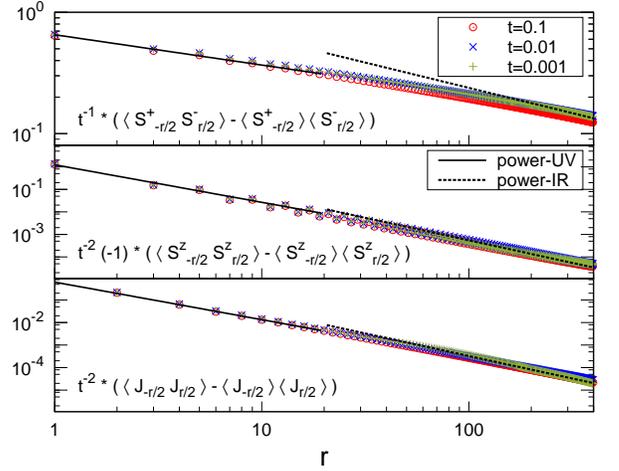}
  \caption{(Color online) Rescaled $\langle S^{\alpha, + }_i S^{\beta, -}_i \rangle$, $\langle S^{\alpha, z }_i S^{\beta, z}_i \rangle$, and $\langle J^{\alpha}_i J^{\beta}_i \rangle$ correlation functions for $g=1.2$. Data for junctions with $t=0.1, 0.01, 0.001$ are plotted. Solid (dotted) lines are power law fitting to the long (short) distance data with IR (UV) exponents from bosonization. (Cf. Table.\ref{table:exponents}.)}
  \label{fig:g>_UV}
\end{figure}

Finally, we  analyze  the UV behavior for the case of $g>1$. Figure~\ref{fig:g>_UV}  plots rescaled correlation functions for the case of $g=1.2$ and various extremely small $t$ in order to expose the UV regime.  In this limit, we are able to fit the numerical results to the boundary perturbation theory results with Dirichlet boundary condition before crossing over  to the long distance behavior. Both the exponents and scaling prefactor agree well. In Fig.~\ref{fig:g>_UV} we also show the power law fitting in the IR limit and crossover from UV to IR exponents are clearly observed.


In summary we present a robust and powerful numerical method to study a junction between two quantum wires using IBC with finite-size DMRG. This method allows, for the first time, to study numerically the crossover of correlation functions near a quantum impurity between the short- and long-distance regimes, as demonstrated by the perfect fit of the UV and IR behaviors between the numerical and bosonization results. This may lead to further exploration of the crossover behavior from UV to IR~\cite{PhysRevB.13.316}.
We also emphasize that this method is also applicable to a more general class of interesting problems, such as the Y-junction for TLL leads~\cite{Oshikawa:2006sw, Rahmani:2010jr,Rahmani:2012bq}, TLL leads with different Luttinger parameters\cite{Hou:2012hz}, and junctions with spin-1/2 interacting fermion leads~\cite{Hou:2008do}.

 This work was supported  by the Ministry of Science and Technology (MOST) of Taiwan under Grants No. 105-2112-M-002-023-MY3, 104-2628-M-007-005-MY3, 104-2112-M-002-022-MY3, and by MEXT/JSPS KAKENHI Grants JP16K05469 and JP17H06462 of Japan. CYL thanks Shuai Yin for helpful discussions. YJK  thanks the hospitality of ISSP, University of Tokyo, where part of the work was done. Numerical calculation was done using Uni10 tensor network library (https://uni10.gitlab.io/) \cite{uni10}.

\bibliography{Ref}

\newpage
\appendix*
\section{Supplemental Material for  Crossover of Correlation Functions near Quantum Impurity in a Tomonaga-Luttinger Liquid}
In this Supplemental Material, we provide additional information on the the construction and the explicit form of the matrix product operators (MPO) and the bosonization derivation of correlation functions.
\date{\today}

\section{Infinite boundary conditions and the effective Hamiltonian}

We follow the approaches used in Refs.~\cite{Michel:2010uo, Phien:2012dr, Phien:2013kh, Milsted:2013iq, Zauner:2015jl} to obtain the effective Hamiltonian for the finite-size window in the MPO form.
We first express the ground state of the bulk in the form of an infinite matrix product state (iMPS).
\begin{align}
  |\Psi\rangle_{\text{bulk}}^{G}& = \sum_{ \{s_i\} }
  \cdots \lambda \Gamma^{s_{i-1}} \lambda \Gamma^{s_i} \lambda \Gamma^{s_{i+1}} \lambda \Gamma^{s_{i+2}} \cdots
  |\mathbf{s}\rangle,\nonumber\\
 &= \sum_{s_i} \cdots A^{s_{i-1}} A^{s_i} \lambda B^{s_i} B^{s_{i+1}} \cdots |\mathbf{s}\rangle.
\end{align}
On the other hand the MPO of the bulk Hamiltonian reads:
\begin{equation}
  H_{\text{bulk}} = \cdots W_{-L-\frac{3}{2}} W_{-L-\frac{1}{2}} \cdots W_{-\frac{1}{2}} W_{\frac{1}{2}} \cdots W_{L+\frac{1}{2}} W_{L+\frac{3}{2}} \cdots,
\end{equation}
where
\begin{equation}
  W_i =
  \left[ \begin{array}{ccccc}
    \mathbbm{1}_i & 0  & 0 & 0 & 0  \\
    S_i^x & 0  & 0 & 0 & 0  \\
    S_i^y & 0  & 0 & 0 & 0  \\
    S_i^z & 0  & 0 & 0 & 0  \\
    0 & -S_i^x  & -S_i^y & \Delta S_i^z & \mathbbm{1}_i
  \end{array} \right].
\end{equation}
We assume that the ground state of the junction Hamiltonian has the following iMPS form:
\begin{equation}
  |\Psi\rangle_{\text{junc}}^{G, 2L} =
  \sum_{s_i} \cdots A^{s_{-L-1}} \left ] \right [ M^{s_{-L}} \cdots M^{s_L} \left ] \right [ B^{s_{L+1}} \cdots |\mathbf{s}\rangle,
\end{equation}
where $L$ is the size of the window. The MPO of the junction Hamiltonian reads:
\begin{equation}
  H_{\text{junc}} = \cdots W_{-L-\frac{3}{2}} W_{-L-\frac{1}{2}} \cdots \widetilde{W}_{-\frac{1}{2}} W_{\frac{1}{2}} \cdots W_{L+\frac{1}{2}} W_{L+\frac{3}{2}} \cdots,
\end{equation}
which is obtained from $H_{\text{bulk}}$ by replacing $W_{-1/2}$ with $\widetilde{W}_{-1/2}$ to represent the link, where
\begin{equation}
  \widetilde{W}_{-1/2} =
  \left[ \begin{array}{ccccc}
    \mathbbm{1}_{-\frac{1}{2}} & 0  & 0 & 0 & 0  \\
    S_{-1/2}^x & 0  & 0 & 0 & 0  \\
    S_{-1/2}^y & 0  & 0 & 0 & 0  \\
    S_{-1/2}^z & 0  & 0 & 0 & 0  \\
    0 & -t S_{-1/2}^x  & -t S_{-1/2}^y & 0 & \mathbbm{1}_{-1/2}
  \end{array} \right].
\end{equation}

\begin{figure}[h]
  \includegraphics[width=0.9\columnwidth]{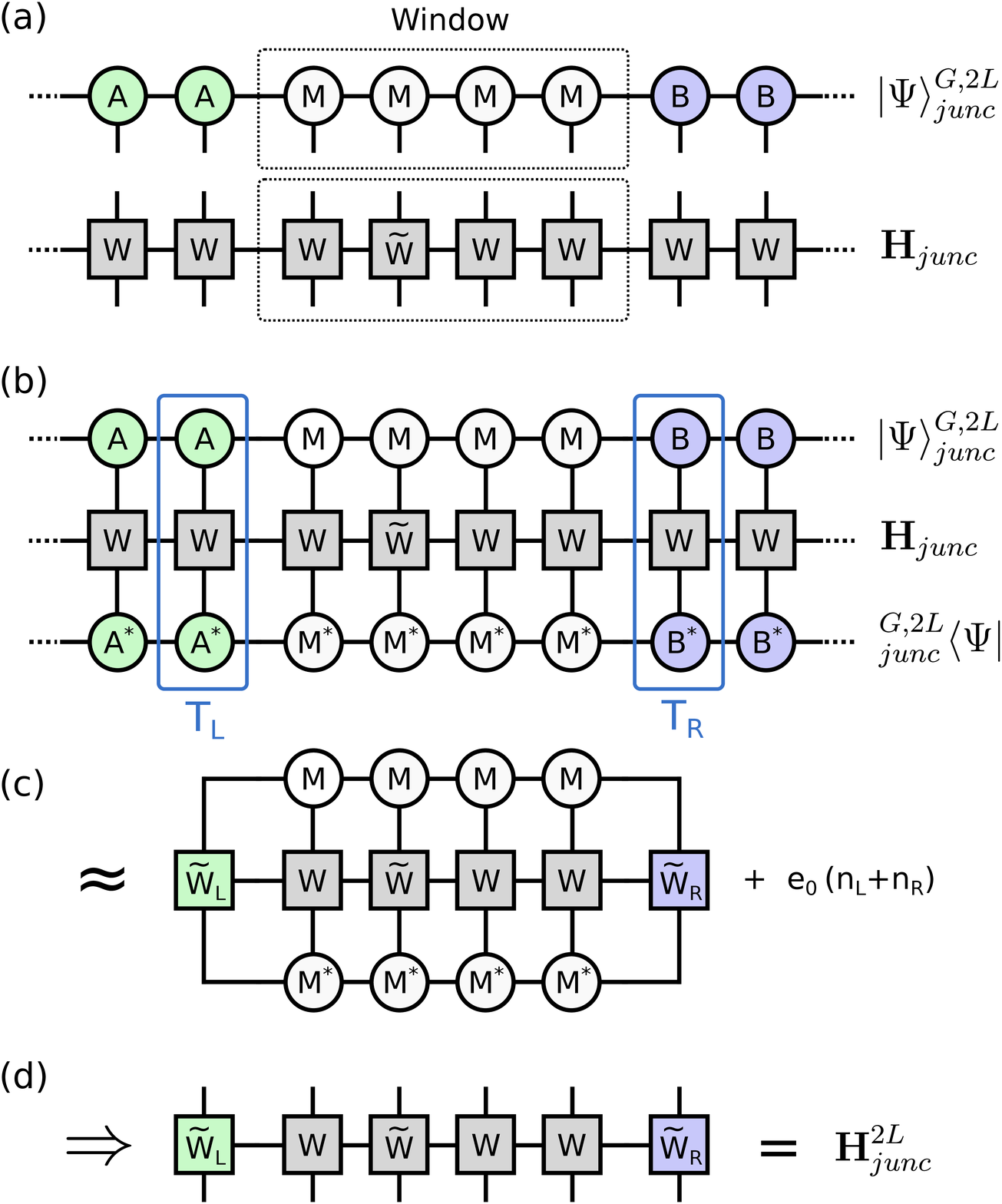}
  \caption{(Color online) Illustration of the process to find the effective Hamiltonian of a finite window.
	(a) Diagrams for the ground state iMPS $|\Psi\rangle^{G,2L}_{junc}$ and the junction Hamiltonian $H_{junc}$.
	(b) Diagram for the expectation value $\langle \Psi|H_{junc}|\Psi \rangle^{G,2L}_{junc}$. (Framed) Left and right generalized transfer matrices.
	(c) Removing the total energy outside the window and attaching IBCs $\widetilde{W}_{\mathcal{L}}$ and  $\widetilde{W}_{\mathcal{R}}$ to represent the left and right semi-infinite extensions respectively.
	(d) The MPO of the effective Hamiltonian $H_{\text{junc}}^{2L}$.}
  \label{fig:h_window}
\end{figure}

In Fig.\ref{fig:h_window}(a) we sketch the tensor network diagram for $|\Psi\rangle_{\text{junc}}^{G, 2L}$ and $H_{\text{junc}}$, while in Fig.\ref{fig:h_window}(b) we sketch the tensor network diagram for $\langle \Psi | H_{\text{junc}} |\Psi\rangle_{\text{junc}}^{G, 2L}$. Since in the thermodynamics limit the total energy $\langle \Psi | H_{\text{junc}} |\Psi\rangle_{\text{junc}}^{G, 2L}$ is divergent, one has to remove the total energy of the sites outside the windows. This results in
\begin{equation}
   \langle \Psi | H_{\text{junc}} |\Psi\rangle_{\text{junc}}^{G, 2L} \sim  \langle \Psi | H_{\text{junc}}^{2L} |\Psi\rangle_{\text{junc}}^{G, 2L} + e_0 (n_L+n_R),
\end{equation}
where $e_0$ is the energy per site of the infinite chain. This leads to the tensor network diagram for $\langle \Psi | H_{\text{junc}}^{2L} |\Psi\rangle_{\text{junc}}^{G, 2L}$ as sketched in Fig.\ref{fig:h_window}(c) and the MPO form of the $H_{\text{junc}}^{2L}$
\begin{equation}
  H_{\text{junc}}^{2L} = \widetilde{W}_{\mathcal{L}} W_{-L-\frac{1}{2}} \cdots \widetilde{W}_{-\frac{1}{2}} W_{\frac{1}{2}} \cdots W_{L+\frac{1}{2}} \widetilde{W}_{\mathcal{R}},
\end{equation}
as sketched in Fig.\ref{fig:h_window}(d). Here $\widetilde{W}_{\mathcal{L}}$ and $\widetilde{W}_{\mathcal{R}}$ are the left and right IBC respectively.

\begin{figure}[h]
  \includegraphics[width=0.9\columnwidth]{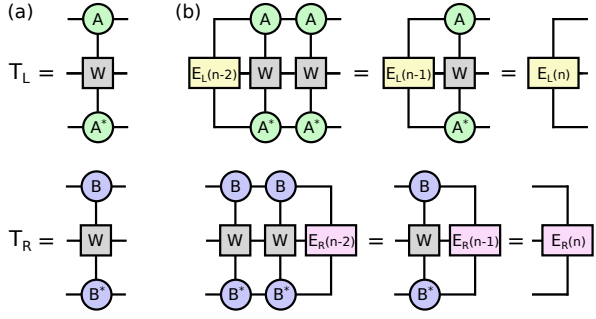}
  \caption{(Color online)(a) Generalized transfer matrices $T_{\mathcal{L, R}}$. (b) Recursion relations of the generalized eigenvectors.}
  \label{fig:eig_vec}
\end{figure}

\begin{figure}[h]
  \includegraphics[width=0.9\columnwidth]{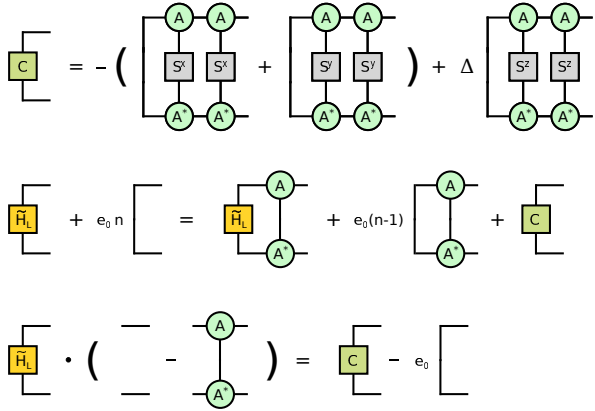}
  \caption{(Color online) The linear equation to solve for $\widetilde{H}_{L}$. $\widetilde{H}_{R}$ is solved in a similar manner, where the tensors are contracted from the right.}
  \label{fig:hb_eqn}
\end{figure}

To identify $\widetilde{W}_{\mathcal{L, R}}$ we consider the generalized transfer matrices $T_{\mathcal{L, R}}$ as sketched in Fig.\ref{fig:eig_vec}(a). Due to the lower triangle structure of $\mathcal{W}$, the transfer matrices $T_{\mathcal{L, R}}$ are not diagonalizable. However, one can use the recursion relations as sketched in Fig.\ref{fig:eig_vec}(b) to show that the left and right generalized eigenvectors $E_{L}(n) $ and $E_R(n)$ have the form
\begin{equation}
  E_L(n) =
  \left[ \begin{array}{ccccc}
    \widetilde{H}_L + e_0 n \widetilde{\mathbbm{1}}_L & -\widetilde{S}_L^x  & -\widetilde{S}_L^y & \Delta\widetilde{S}_L^z & \widetilde{\mathbbm{1}}_L
  \end{array} \right],
\end{equation}
\begin{equation}
  E_R(n) =
  \left[ \begin{array}{c}
    \widetilde{\mathbbm{1}}_R  \\
    \widetilde{S}_R^x  \\
    \widetilde{S}_R^y   \\
    \widetilde{S}_R^z  \\
    \widetilde{H}_R + e_0 n \widetilde{\mathbbm{1}}_R
  \end{array} \right],
\end{equation}
where
\begin{equation}
  \widetilde{S}^{x,y,z}_L = \sum_{ss^\prime} \langle s |S^{x,y,z}| s^\prime \rangle A^{\dagger s^\prime} A^s,
\end{equation}
and
\begin{equation}
  \widetilde{S}^{x,y,z}_R = \sum_{ss^\prime} \langle s |S^{x,y,z}| s^\prime \rangle B^s B^{\dagger s^\prime}.
\end{equation}
Furthermore $\widetilde{H}_{L,R}$ is obtained by solving a linear equation as sketched in Fig.\ref{fig:hb_eqn}. Finally we obtain the left and right IBC $\widetilde{W}_{L,R}$ by dropping the the divergent energy of the semi-infinite extension outside the window to get

\begin{equation}
  \widetilde{W}_L =
  \left[ \begin{array}{ccccc}
    \widetilde{H}_L & -\widetilde{S}_L^x  & -\widetilde{S}_L^y & \Delta\widetilde{S}_L^z & \widetilde{\mathbbm{1}}_L
  \end{array} \right] \text{and }
  \widetilde{W}_R =
  \left[ \begin{array}{c}
    \widetilde{\mathbbm{1}}_R  \\
    \widetilde{S}_R^x  \\
    \widetilde{S}_R^y   \\
    \widetilde{S}_R^z  \\
    \widetilde{H}_R
  \end{array} \right].
\end{equation}

\section{Bosonization}

In this section, we explain the detail of the analytical calculation we have done in the main text.

First we used the bosonization for the spin chain by Lukyanov\cite{Lukyanov:2002fg},
\begin{align}
 S^{-}_j & \sim   e^{- i \theta }\left(b+c\left(-1\right)^j\mbox{sin}\left(2\phi\right)\right) , \\
 S^z_j & \sim \frac{1}{\pi }\frac{\partial\phi}{\partial x}+a(-1)^j \sin{2\phi} , \\
J_r & =i\left(S^{+}_{j+1}S^{-}_{j}-S^{+}_{j}S^{-}_{j+1}\right)\sim \frac{gv}{\pi} \frac{\partial \theta}{\partial x}.
\end{align}
Here $g$ is the Luttinger parameter and $v$ is the spin-wave velocity.
The Lagrangian density and the definition of the bosonic field are,
\begin{equation}
 \mathcal{L} = \frac{1}{2\pi g} (\partial_\mu \phi)^2 .
\label{eq:Lag}
\end{equation}
\begin{align}
 \langle \phi_R(0) \phi_R(z) \rangle & \sim
  - \frac{g}{4} \log{z} + \mbox{const.} , \\
 \langle \phi_L(0) \phi_L(\bar{z}) \rangle & \sim
  - \frac{g}{4} \log{\bar{z}} + \mbox{const.}. \\
\end{align}
We used the complex coordinate $z=x+it$ and assumed
the geometry of the theory without junction is a full complex plane.

The field $\phi$ is subject to the compactification
\begin{equation}
 \phi = \phi + \pi .
\end{equation}
Using this, the dual field $\theta$ is defined as
\begin{equation}
\theta \equiv (1/g) ( \phi_L - \phi_R ) ,
\end{equation}
where $\theta$ is subject to the compactification
\begin{equation}
\theta \sim \theta +2\pi.
\end{equation}

Then we used the boundary perturbation procedure.
For Dirichlet boundary condition, we gradually introduce
interaction to two decoupled chains through the boundary.
The geometry is two half plane with the interaction on the line $x=0$.
We note the calculation for $S_{r}^{\alpha +}S_{r}^{\beta -}$ as an example.
The first order perturbation leads to,
\begin{equation}
 t \int d\tau \;
\left[ \frac{1}{2}
\langle S^{+\alpha}_{i+1/2}(0) S^{-\alpha}_{1/2}(\tau)\rangle_{D}
\langle S^{+\beta }_{1/2}(\tau) S^{-\beta}_{i+1/2}(0) \rangle_{D}
\right],
\label{eq.SpSm_1st}
\end{equation}
The Dirichlet boundary condition for bosonic field is,
\begin{equation}
\phi^L_\gamma(0,\tau)  + \phi^R_\gamma(0,\tau) = 0, \gamma=\alpha, \beta.
\end{equation}
This can be resolved by an analytical continuation of $\phi^R_2$ to
$x<0$ and
\begin{equation}
 \phi^L_\gamma(x,\tau) \equiv - \phi^R_\gamma(-x,\tau).
\end{equation}
Using this, the correlation function is given as
\begin{align}
& t \int d\tau \;
\left(\frac{1}{r^2+\tau^2}\right)^{1/g} (2r)^{1/2g}  \\
\sim &
t r^{-(2/g-1)} r^{1/2g} = t r^{-(3/2g-1)} .
\end{align}
For free boundary condition, we think of the junction
as defect in CFT.
We evaluated this defect by using operator product expansion
for CFT.
\begin{equation}
\begin{split}
H_{\text{link}} = (1-t) ( S^{+\alpha}_{1/2} S^{-\beta}_{1/2} +S^{-\alpha}_{1/2} S^{+\beta}_{1/2} ) \\
\sim \mu_{B}\int ^{\infty}_{-\infty}d\tau \mbox{cos}\left( 4\phi (i\tau) \right).
\end{split}
\end{equation}
We have introduced the parameter $\mu_B$ for the boundary perturbation parameter.
For Dirichlet boundary condition, it is proportional to $t$.
For the connected wire, it is proportional to $1-t$.

\section{Bulk perturbation effect}
In this section, we show the results of the bulk perturbation.
The field theoretic bulk action for XXZ spin chain is written as
\begin{equation}
S=S_{CFT}+\mu\int dx'd\tau '\cos \left( 4\phi\right).
\end{equation}
By adding the contribution of this term, we can obtain the results
in the main section.
The $\mu$ is determined by $V$ and it is explicitly determined without a
junction.

For example, we show the correction of the results in the previous section.
For Dirichlet boundary condition, the stagger part of the field gives the following
correction of the term for each wire,
\begin{equation}
\langle S^{+\alpha}_{r+1/2}(0)S^{-\alpha}_{1/2}(\tau)\rangle_{\rm uniform}=r^{-\frac{3}{4g}}C'_{0}+\mu r^{-4g-\frac{3}{4g}+2}C'_{1},
\end{equation}
where $C'_0(\tau/r)$ and $C'_1(\tau/r)$ are functions of $\tau/r$.
The second term is obtained from the following integral,
\begin{equation}
\begin{split}
&\mu \int dx' d\tau ' b^2 e^{i\theta (r) R} e^{-i\theta (i\tau) R} \cos \left(  4\phi(x', \tau ')\right) \\
&=\mu r^{-4g-\frac{3}{4g}+2}C'_{1} \left(\frac{\tau}{r}\right).
\end{split}
\end{equation}
The stagger term of the field gives,
\begin{equation}
\langle S^{+\alpha}_{i+1/2}(0)S^{-\alpha}_{1/2}(\tau)\rangle_{\rm stagger}=r^{-\frac{3}{4g}-g}C_{0}+\mu r^{-5g-\frac{3}{4g}+2}C_{1}.
\end{equation}
Then by composing contribution of each wire and the integration of
$\tau$, we can obtain the desired correlation function.
However, we do not know the effect of $(\textrm{stagger})\times(\textrm{uniform})$ terms and the precise prefactor of them.
The results for $(\textrm{stagger})\times(\textrm{stagger})$ are,
\begin{align}
\langle S^{+ \alpha}_{r+1/2} S^{- \beta}_{r+1/2} & \rangle=\left( -1\right)^{r}\mu_B \nonumber \\ 
&\left( r^{-\frac{3}{2g}-2g+1}
+\mu r^{-\frac{3}{2g}-6g+3}+
\mu^{2} r^{-\frac{3}{2g}-10g+5}\right),
\end{align}
\begin{align}
\langle S^{z \alpha}_{r+1/2} S^{z \beta}_{r+1/2} &\rangle=\left( -1 \right)^{r} \mu_{B}^{2} \nonumber \\
&\left( r^{-\frac{2}{g}-2g+2}+\mu r^{-\frac{2}{g}-6g+4}+\mu^{2} r^{-\frac{2}{g}-10g+6} \right).
\end{align}
The results for $(\textrm{uniform})\times(\textrm{uniform})$ are,
\begin{align}
\begin{split}
\langle S^{+ \alpha}_{r+1/2} S^{- \beta}_{r+1/2}\rangle=\mu_{B}
\left( r^{-\frac{3}{2g}+1}+\mu r^{-\frac{3}{2g}-4g+3}+\mu^2 r^{-\frac{3}{2g}-8g+5} \right),
\end{split}
 \\
\begin{split}
\langle S^{z \alpha}_{r+1/2} S^{z \beta}_{r+1/2}\rangle=\mu^{2}_{B}
\left( r^{-\frac{2}{g}}+\mu r^{-\frac{2}{g}-4g+2}+\mu^{2}r^{-\frac{2}{g}-8g+4}\right),
\end{split}
\\
\begin{split}
\langle J^{\alpha}_{r+1/2} J^{\beta}_{r+1/2}\rangle=\mu_{B}^{2} \left( r^{-\frac{2}{g}}+\mu r^{-\frac{2}{g}-4g+2} +\mu^{2}r^{-\frac{2}{g}-8g+4}\right).
\end{split}
\end{align}
Here and the following discussion we omit the constants for simplicity. 

For the free boundary condition, we can get the correction by the same procedure, but
it does not cause the problem of the mixing of stagger and uniform terms.
The lowest order of the boundary perturbation can be changed by the effect of the bulk perturbation
in this case.
The results are
\begin{align}
\begin{split}
\langle S^{+\alpha}_{r+1/2} S^{-\beta}_{r+1/2} \rangle&=r^{-\frac{1}{2g}}+\mu^{2}r^{-\frac{1}{2g}-8g+4}+\mu\mu_{B}r^{-\frac{1}{2g}-8g+3} \\
&+\mu^{2}_{B}r^{-\frac{1}{2g}-8g+2}+\left(-1\right)^{r}r^{-\frac{1}{2g}-2g} \\
&+\left(-1\right)^{r}\mu r^{-\frac{1}{2g}-6g+2}+\left(-1\right)^{r}\mu_{B}r^{-\frac{1}{2g}-6g+1}.
\end{split}
\\
\begin{split}
\langle S^{z\alpha}_{r+1/2} S^{z\beta}_{r+1/2} \rangle&=r^{-2}+\mu^{2}r^{-8g+2}+\mu\mu_{B}r^{-8g+1} \\
&+\mu^{2}_{B}r^{-8g}+\left(-1\right)^{r}r^{-2g} \\
&+\left(-1\right)^{r}\mu r^{-6g+2}+\left(-1\right)^{r}\mu_{B}r^{-6g+1}.
\end{split}
\\
\begin{split}
\langle J_{r+1/2} J_{r+1/2} \rangle&=r^{-2}+\mu^{2}r^{-8g+2}+\mu\mu_{B}r^{-8g+1} \\
&+\mu^{2}_{B}r^{-8g}.
\end{split}
\end{align}
In any case, all of subleading terms resulted from the correction of the bulk perturbation
 decay faster than leading order terms we as have explained in
the previous section. Hence that verifies the validity of the results in the main text.

\end{document}